\documentclass[twocolumn,showpacs,preprintnumbers,amsmath,amssymb]{revtex4-1}

\usepackage{cmap}
\usepackage{color}
\usepackage{latexsym,amsmath,amsbsy,graphicx}
\usepackage{float}
\usepackage{footnote}

\usepackage{multirow}
\usepackage{epstopdf}
\usepackage{graphicx}
\usepackage{dcolumn}
\usepackage{bm}
\usepackage{color}
\graphicspath{{Img/}}
\DeclareGraphicsExtensions{.eps}

\newcommand{\AaAa}{{$A$+$A$\ }}

\newcommand{\PbPb}{{$Pb$+$Pb$\ }}

\newcommand{\hydjet}{{\sc hydjet++ }}
\newcommand{\pyquen}{{\sc pyquen }}
\newcommand{\fastmc}{{\sc fastmc }}
\newcommand{\share}{{\sc share }}
\newcommand{\ampt}{{\sc ampt }}
\newcommand{\hijing}{{\sc hijing }}
\newcommand{\pythia}{{\sc pythia }}

\newcommand{\jetset}{{\sc jetset }}
\newcommand{\art}{{\sc art }}
\newcommand{\zpc}{{\sc zpc }}

\newcommand{\sqrts}{\sqrt{s}}

\newcommand{\ds}{{\displaystyle}}

\begin{document}   

\title{Modeling net-charge fluctuations in heavy-ion collisions at the LHC}

\author{G.O.~Ambaryan$^1$, A.S.~Chernyshov$^1$, G.Kh.~Eyyubova$^2$, 
V.L.~Korotkikh$^2$, I.P.~Lokhtin$^2$, S.V.~Petrushanko$^2$, 
A.M.~Snigirev$^{2,3}$, and E.E.~Zabrodin$^{2,4}$}
	

\affiliation{$^1$ Physics Department, Lomonosov Moscow State University, 
RU-119991 Moscow, Russia }
\affiliation{$^2$ Skobeltsyn Institute of Nuclear Physics, Lomonosov 
Moscow State University, RU-119991 Moscow, Russia }
\affiliation{$^3$ Bogoliubov Laboratory of Theoretical Physics, Joint 
Institute for Nuclear Research, RU-141980 Dubna, Russia }
\affiliation{$^4$ Department of Physics, University of Oslo, 
N-0316 Oslo, Norway}

\begin{abstract}
Analysis of \PbPb data for net-charge fluctuations at LHC energies 
using the \hydjet model is presented. The strongly intensive
quantities $D$ and $\Sigma$ were used to remove the effects related 
to system volume fluctuations. We employed two versions of 
\hydjet for the analysis. The first one is the standard or default 
version, whereas the second one is a modification that takes into 
account explicit event-by-event conservation of the electric 
net-charge of produced particles. The inclusion of the canonical 
net-charge conservation in the model allows for better description 
of the experimental data obtained by the ALICE and CMS 
Collaborations. A comparison with calculations from other models 
is also presented.

\end{abstract}


\maketitle

\section{ Introduction }
\label{intro}

One of the current problems of modern high-energy physics is the study 
of strong interactions at extremely high temperatures and energy 
densities reached in relativistic heavy-ion collisions (for a recent 
review, see, for instance \cite{Harris:2023tti} and references therein). 
This includes the investigation of properties of created quark-gluon 
plasma (QGP) and dynamics of quark-hadron phase transition, as
well as the analysis of the mechanisms of multiparticle production. 
The particular interest of this subject is conditioned by the intensive 
experimental studies conducted at the Large Hadron Collider (LHC). The 
maximum energies currently achieved in laboratory conditions at the LHC 
allow for probing the high-temperature QGP state, whose properties are
similar to those of ``proto-matter'' in Early Universe. Future 
experiments with heavy-ion beams at intermediate energies in 
Nuclotron-based Ion Collider fAcility (NICA) and Facility for 
Antiproton and Ion Research (FAIR) are expected to have a significant 
impact on the study of dynamics of the quark-hadron phase transition, 
including search for the ``critical point'', near their boundary. Both 
aforementioned tasks are complementary.

Among various physical observables, event-by-event (EbyE) fluctuations 
of conserved quantities such as net-baryon, net-strangeness, and 
net-electric charge are considered sensitive tools to characterize 
the thermodynamic properties and critical behavior of hot and dense 
matter produced in relativistic heavy-ion collisions 
\cite{Asakawa:2000wh,Jeon:2003gk}. In particular, the net-charge 
fluctuations are proportional to the square of the charges in the 
system, and may be quantified in terms of $D$ 
\cite{Jeon:2000wg,Pruneau:2002yf} and $\Sigma$ 
\cite{Gorenstein:2011vq,Gazdzicki:2013ana,Wu:2021vix} variables. 

A dramatic decrease of the EbyE fluctuations of the net charges in 
local regions of a phase space was predicted by many theoretical 
investigations \cite{Asakawa:2000wh,Jeon:2000wg,BJK00,HJ01}
as a prominent signal of QGP formation, provided that the 
final-state charge fluctuations survive the hadronization. Note 
that these fluctuations are related to the charge distribution in 
the plasma state and not to the phase transition from plasma to 
hadrons. The main question is how and why the initial distributions 
can survive during the QGP hadronization stage \cite{SS01,FW01}.
 
Experimental studies of net-charge fluctuations in heavy-ion 
collisions, including measurements of their dependencies on rapidity 
interval and event centrality, were conducted by the NA49 
Collaboration at CERN Super Proton Synchrotron (SPS) 
\cite{na49_1,na49_2}, STAR Collaboration at BNL Relativistic Heavy 
Ion Collider (RHIC) \cite{STAR:2008szd}, and 
ALICE~\cite{ALICE:2012xn,Khan:2022ndt} 
and CMS~\cite{hin-22-005} Collaborations at CERN LHC. The 
corresponding experimental data are poorly described by existing 
theoretical models.

In the present study, the phenomenological analysis of \PbPb data for 
net-charge fluctuations at LHC energies was performed in terms of the 
strongly intensive quantities $D$ and $\Sigma$. The advantage of 
using these strongly intensive quantities is that they are independent 
of both system volume and fluctuations 
\cite{Gorenstein:2011vq,Gazdzicki:2013ana}. 
The paper is organized as follows. To simulate heavy-ion
collisions, we selected the two-component Monte Carlo \hydjet 
model~\cite{Lokhtin:2008xi}. Their final state represents a 
superposition of the soft ``thermal'' and hard states resulting 
from multi-parton fragmentation. Recently 
\cite{Chernyshov:2022oik}, we presented a modification of this model 
through the inclusion of explicit event-by-event charge conservation 
using a statistical approach. This modification allows for 
describing the experimentally observed centrality dependence of 
widths of the charge balance function in \PbPb collisions at the LHC 
center-of-mass energy of 2.76~TeV per nucleon pair. Basic features of 
the default version of \hydjet and its modification for exact EbyE 
charge conservation are presented in Section~\ref{sec_2}. 
Section~\ref{sec_3} discusses the relation between the charge 
balance function and net-charge fluctuations, first established 
in~\cite{Pruneau:2002yf}. 
In Section~\ref{sec_4}, the results obtained for the net-charge 
fluctuations in \PbPb collisions at a center-of-mass energy of 
5.02~TeV per nucleon pair using the \hydjet model are compared with 
ALICE data~\cite{Khan:2022ndt} and with the predictions of other 
Monte Carlo models, such as 
\hijing~\cite{Gyulassy:1994ew,hijing2,hijing3} 
and \ampt~\cite{Zhang:1999bd,ampt2,ampt3}.
Conclusions are drawn in Section~\ref{sec_5}. 

\section{ Basic principles of employed models }	
\label{sec_2}
\subsection{\hydjet (default version)}
\label{subsec_2.1}

The \hydjet model (Hydrodynamics with JETs) was designed as a Monte
Carlo (MC) event generator \cite{Lokhtin:2008xi,hydjet_2} for 
simulation of heavy-ion collisions at (ultra)relativistic energies. 
It consists of two parts describing the soft and hard processes, 
respectively. The soft sector of \hydjet is the adapted version 
of the \fastmc model \cite{fastmc_1,fastmc_2} and represents a 
relativistic hydrodynamic parametrization of hypersurfaces of chemical 
and thermal freeze-out under given freeze-out conditions. Although the 
model includes the option that both freeze-outs can occur 
simultaneously in a single freeze-out scenario, much better agreement 
with experimental data is obtained in a scenario with separated 
chemical and thermal freeze-outs. The simulation of a single event 
begins by calculating the effective fireball volume $V_{eff}$, which 
depends on the mean number of participating nucleons for a given 
impact parameter $b$, i.e., the collision centrality. This impact 
parameter is determined from the Glauber multiple scattering model. 
For the most appropriate scheme with separate freeze-outs, the 
composition of particles in the system is frozen at the chemical 
freeze-out stage according to the predictions of statistical thermal 
models \cite{sm1,sm2}. The system continues to expand and cools to the 
point of thermal freeze-out, where the mean free path of hadrons 
exceeds the size of the system. At this stage, two- and many-body 
decays of resonances are taken into account as final state interactions 
(FSIs). The table of resonances included in \hydjet from the model 
\share \cite{share} is notably rich. It contains more than 360 mesonic 
and baryonic states, including the charmed ones. To manage the decays 
of resonances, \hydjet employs its own original routine.

The hard sector of the model deals with the propagation of hard partons 
through an expanding quark-gluon plasma. The approach, which takes into 
account both gluon radiation due to parton rescattering and partonic 
collisional losses, is based on the \pyquen (PYTHIA QUENched) model of 
parton energy losses \cite{pyquen}. The number of jets is generated in
accordance with a binomial distribution, whereas their average number 
in a nuclear collision with a given impact parameter $b$ is calculated 
as a product of the number of binary nucleon-nucleon ($NN$) collisions 
and integral cross-section of the hard process in $NN$ collisions
with the minimum transverse momentum transfer $p_T^{min}$. This is 
an important free parameter of the \hydjet model, because partons 
created in (semi)hard subprocesses with momentum transfer below 
$p_T^{min}$ are considered to be in thermal equilibrium. The 
products of their hadronization are added to the spectra of soft 
particles.

The simultaneous consideration of soft and hard processes allows 
\hydjet to describe aspects such as centrality and transverse 
momentum dependence of the differential elliptic and triangular 
flow of hadrons in heavy-ion collisions at energies of RHIC and LHC 
\cite{Eyyubova:2009hh,Bravina:2013xla,prc_17}, 
violation of the mass hierarchy and violation of the constituent 
quark scaling (NCQ) of these differential flows at intermediate and 
high values of $p_T$ \cite{jpg_10,Bravina:2013xla,prc_14}, 
peculiarities of factorization of dihadron angular harmonics (ridge)
\cite{Eyyubova:2014dha}, and higher flow harmonics 
\cite{Bravina:2013xla,prc_14} and their EbyE fluctuations 
\cite{Bravina:2015sda}. However, to reproduce both the centrality 
dependence and the width of the balance functions of charged 
particles, the model requires modification.

\subsection{ Modified \hydjet }
\label{subsec_2.2}

The generation of hadron species in the soft sector of \hydjet 
proceeds within the framework of the grand canonical ensemble (GCE), 
similar to other statistical models of hadron-resonance gas. 
Therefore, the correlations between hadrons arise because of decays 
of resonances and jet fragmentation. These correlation mechanisms 
appear insufficient to describe features of the balance functions of 
charged hadrons in \PbPb collisions at LHC energies. To reproduce 
this signal, the model was extended in \cite{Chernyshov:2022oik}. The 
modified version of \hydjet takes into account exact conservation of 
the net electric charge in the system of colliding nuclei on the EbyE 
basis. 

Next, we describe the algorithm for the hadronic system with zero
net electric charge. All hadrons of the soft component are 
generated according to the GCE prescription, i.e., unlike-sign 
charges are initially uncorrelated. Then, half of the charged hadrons
are randomly discarded. A Monte Carlo realization of this procedure 
assumes that each directly produced charged hadron is either removed 
with probability 50\% from the particle spectrum or remained untouched. 
Then, for each of the remaining hadrons, its counterpart with opposite 
electric charge and similar transverse momentum is generated. As 
reported in \cite{Chernyshov:2022oik}, this procedure does not change 
the pseudorapidity, angular, and transverse momentum spectra of charged 
particles for statistics of at least 1000 events.
The azimuthal angle $\phi_2$ and rapidity $\eta_2$ of newly produced 
particles are distributed around the azimuthal angle $\phi_1$ and 
rapidity $\eta_1$ of the initial hadron following a Gauss normal 
distribution,
\begin{equation}
\ds
P_{\mu,\sigma}(x) = \frac{1}{\sqrt{2 \pi} \sigma} \exp{\left[ 
   - \frac{(x - \mu)^2}{2 \sigma^2} \right]} \ ,
\label{gauss}
\end{equation}
where $x = (\phi_2,\eta_2)$, $\mu = (\phi_1,\eta_1)$, and 
$\sigma = (\sigma_\phi, \sigma_\eta)$. Both dispersions, 
$\sigma_\phi$ and $\sigma_\eta$, characterize the strength of the 
charge correlations of the produced hadrons. Their values are free 
parameters of the modified model and should be fixed by comparison 
with the available experimental data on the balance functions. The 
described procedure is relevant for the midrapidity region of 
heavy-ion collisions at RHIC and LHC energies. For lower energies, 
this procedure was recently generalized to systems with non-zero 
net electric charge \cite{Chern_jetp_2024}.  

\subsection{ \hijing and \ampt models }
\label{subsec_2.3}

Concerning \hijing (Heavy Ion Jet INteraction Generator) 
\cite{Gyulassy:1994ew,hijing2,hijing3}, it is a Monte Carlo 
event generator designed for description of jet and mini-jet 
production and associated particle production in high energy 
hadronic ($hh$), hadron-nucleus ($hA$), and nuclear (\AaAa) 
collisions. Recall that jets with large transverse momenta are one 
of the hard probes of heated nuclear matter. The sources of these 
jets are particles that arise during the hadronization of quarks 
and gluons. Before hadronization, quarks and gluons propagate in 
the nuclear medium, which can include both hadrons and a quark 
component. In this case, the energy loss of a quark during 
collisions with particles of the medium and during bremsstrahlung 
depends on the properties of the medium. This phenomenon, known as 
``jet quenching", can serve as a signature of quark-gluon plasma. 
Another important process is the formation of mini-jets. 
It is estimated that mini-jets contribute up to 50\% of the 
transverse energy in central collisions of heavy ions. Mini-jets 
with medium transverse momenta have shorter path lengths in the 
parton medium and are more likely to dissipate in the background, 
losing memory of the initial correlations and changing their 
characteristics when interacting with the medium.
Thus, \hijing incorporates several mechanisms such as multiple 
mini-jet production, soft excitation, nuclear shadowing of parton 
distribution functions, and jet interactions in dense hadronic 
matter.

To describe hadron production in relativistic collisions, \hijing 
relies on a two-component model developed initially for $hh$
interactions \cite{GaHa85,DuHo87,SjZi87,ChHw89} and extended to 
\AaAa collisions \cite{BlMu87,KLL87,EKL89}. This two-component 
model treats the parton interactions in high-energy $hh$ collisions 
as (semi-)hard with jet production if the transverse momentum of 
the jet is larger than the characteristic cut-off parameter $p_0$, 
and as soft in the opposite case. To generate kinematic variables 
for each hard scattering, \hijing uses subroutines of \pythia 
\cite{SjZi87,pythia_2}, whereas for string fragmentation \jetset 
routine \cite{jetset} of the Lund model \cite{lund1,lund2} is 
employed. Further details of the model can be found in 
\cite{Gyulassy:1994ew,hijing2,hijing3}.

Regardibg \ampt (A Multi-Phase Transport) 
\cite{Zhang:1999bd,ampt2,ampt3}, it is a hybrid transport model 
designed for the description of $pA$ and \AaAa collisions at 
relativistic and ultra-relativistic energies. It consists of four 
main parts. To generate the initial conditions, such as production 
of hard mini-jet partons and soft strings, the model employs 
results from \hijing. The further rescattering of partons is 
governed by Zhang's parton cascade (\zpc) \cite{Zh98}, which 
manages the two-body parton interactions within the framework of 
perturbative quantum chromodynamics (pQCD). Two scenarios are used 
to hadronize partons when they stop interacting. In the first 
scenario, partons are recombined with parent strings, which 
convert into hadrons according to the Lund fragmentation model 
\cite{lund1,lund2}. The second scenario assumes the melting of 
strings into partons at the initial stage and hadronization of 
partons at the end of the parton cascade within the quark 
coalescence model. Finally, hadronic matter propagates and 
develops a hadron cascade described by the \art (A Relativistic 
Cascade) model \cite{art1,art2}.

In our calculations, we used the \ampt model with string melting.
Other parameters of the model were extracted from \cite{ZWL14_prc}. 

\section{ Net-charge fluctuations and charge balance function }	
\label{sec_3}

In relativistic heavy-ion collisions, one can denote the net 
charge as $Q = N_+ - N_-$, the total number of charged particles as 
$N_{ch} = N_+ + N_-$, and the ratio of positive to negative charges
as $R = \frac{N_+}{N_-}$. Then, the following relation between 
the variances $\langle \delta R^2 \rangle$ and $\langle \delta Q^2 
\rangle$ can be established, provided that $\langle N_{ch} \rangle 
\gg \langle Q \rangle$ \cite{Jeon:2000wg}:

\begin{equation}	
\ds
\label{eq:1}
\langle N_{ch}\rangle \langle \delta R^2 \rangle = D =
     4 \frac{\langle \delta Q^2 \rangle}{\langle N_{ch} \rangle} \ .
\end{equation}

Here, $D$ is the final observable measure of fluctuations 
\cite{Asakawa:2000wh,Jeon:2000wg}.
This measure can be linked to the event-by-event fluctuations
by calculating the second moment, which is the difference between 
the relative charged particle multiplicities 
$\nu_{+-}$ and statistical fluctuations $\nu_{+-,\text{stat}}$

\begin{eqnarray}
\ds	
\label{eq:3}
\nu_{+-,\text{dyn}} &=& \nu_{+-} - \nu_{+-,\text{stat}} \ , \\ 
\label{eq:4}
\nu_{+-} &=& \left\langle \left( \frac{N_+}{\langle N_+ \rangle} -
         \frac{N_-}{\langle N_- \rangle} \right)^2 \right\rangle \ , \\
\label{eq:5}
\nu_{+-,\text{stat}} &=& \frac{1}{\langle N_+ \rangle} + 
                  \frac{1}{\langle N_- \rangle} \ . 
\end{eqnarray}

Therefore, the fluctuation measure $D$ is expressed as 
\cite{Jeon:2000wg}
\begin{equation}
\ds	
\label{eq:6}
D= 4 \frac {\langle \delta Q^2 \rangle} {\langle N_{ch} \rangle} = 
   \langle N_{ch} \rangle \nu_{+-,\text{dyn}} + 4 \ .
\end{equation}

From Eqs.(\ref{eq:3}) and (\ref{eq:4}) we can derive another 
expression for $\nu_{+-,\text{dyn}}$
\begin{equation}
\ds
\label{eq:7}
\nu_{+-,\text{dyn}}=\frac{\langle N_+^2 \rangle - \langle N_+ \rangle}
          {\langle N_+ \rangle^2} + 
   \frac{\langle N_-^2 \rangle - \langle N_- \rangle}
          {\langle N_- \rangle^2} - 2
   \frac{\langle N_+ N_- \rangle}{\langle N_+ \rangle 
                     \langle N_- \rangle} \ ,
\end{equation}
which means that the correlations are negative when the covariance 
term dominates.

Another measure is the strongly intensive quantity (SIQ), suggested 
in \cite{Gorenstein:2011vq} for the analysis of particle multiplicity 
correlations and fluctuations in high-energy \AaAa collisions. It 
characterizes the second moment of random extensive variables used to 
study fluctuations and correlations in a physical system.
  
For two extensive quantities $A$ and $B$ the strongly intensive quantity 
$\Sigma [A,B] $ reads
\begin{equation} 
\ds
\label{eq:9}
\Sigma [A,B] = \frac{1}{C_\Sigma} \bigg[ \langle B \rangle \omega[A] +  
   \langle A \rangle \omega[B] - 2(\langle AB \rangle - 
                 \langle A \rangle \langle B \rangle ) \bigg] \ ,
\end{equation}
where
\begin{equation}
\ds	
\label{eq:10}
\omega[A] = \frac{\langle A^2 \rangle - \langle A \rangle^2}
   {\langle A \rangle}\ , \quad
\omega[B] = \frac{\langle B^2 \rangle - \langle B \rangle^2}
   {\langle B \rangle} \ .
\end{equation}

For particle charged multiplicities $A = N_+$ and $B = N_-$ we have 
$C_\Sigma = \langle N_+ \rangle + \langle N_- \rangle$ and
\begin{equation}
\ds	
\label{eq:11}
\begin{aligned}
\Sigma [N_+, N_-] &= \left\{ \langle N_-\rangle 
\frac{\langle N_+^2\rangle - \langle N_+ \rangle^2}{\langle N_+ \rangle} 
  +   \langle N_+ \rangle \frac{\langle N_-^2\rangle -
            \langle N_- \rangle^2}{\langle N_- \rangle } -\right. \\
  &\bigg. - 2 (\langle N_+ N_- \rangle - \langle N_+ \rangle 
    \langle N_- \rangle ) \bigg\} \bigg/ 
    \bigg( \langle N_+ \rangle + \langle N_- \rangle \bigg) \ .
\end{aligned}
\end{equation}
Comparing this with the result obtained using Eq.(\ref{eq:7}), we have 
that
\begin{equation}
\ds
\label{eq:12}
 \frac{\langle N_+ \rangle + \langle N_- \rangle }{\langle N_+ \rangle 
 \langle N_- \rangle} (\Sigma [N_+, N_-] - 1)  = \nu_{+-,\text{dyn}} \ .
\end{equation}

It is possible to demonstrate that the strongly intensive quantity 
$\Sigma [N_+, N_-]$ is related to the integral of the balance function 
$B(\Delta\eta)$ \cite{Chernyshov:2022oik} through the equation
\begin{equation}
\ds
\label{eq:13}
	\Sigma [N_+, N_-] = 1 - \int B(\Delta\eta)d\Delta\eta \ .
\end{equation}
Here we define the charge balance function as 
\begin{equation}
\ds
\label{eq:14}
B(\Delta\eta) = \frac{1}{2}\left\{\frac{\langle N_{+-} \rangle - 
                \langle N_{++} \rangle}{\langle N_+ \rangle} + 
                \frac{\langle N_{-+}\rangle - \langle N_{--} \rangle}
                     {\langle N_- \rangle}\right\}\ ,
\end{equation}
where $N_{+-} = N_{+-}(\Delta\eta)$ is the number of pairs of 
unlike-sign charged particles separated by relative pseudorapidity 
$\Delta\eta$, and $N_+$ is the number of positively charged particles 
within pseudorapidity window $|\eta| < H$. Other terms in 
Eq.(\ref{eq:14}) are defined similarly. From this point onward, we 
assume that within the rapidity window, $\langle N_+ \rangle = 
\langle N_- \rangle = \frac{\langle N_\text{ch} \rangle}{2}$; thus, 
Eq.(\ref{eq:14}) can be integrated over $\Delta\eta$ and reduced to
\begin{equation}
\label{eq:15}
\int B(\Delta\eta)d\Delta\eta = \frac{1}{\langle N_\text{ch} \rangle}
    \bigg\{2 \langle N_{+-} \rangle - \langle N_{++} \rangle - 
                     \langle N_{--}\rangle\bigg\} \ ,
\end{equation}
where the number of charged particle pairs no longer depends on 
the relative rapidity. Under the same assumption of charge balance, 
the term $\Sigma[N_+, N_-] - 1$ in Eq.(\ref{eq:11}) can be rewritten 
using relations $\langle N_{+-} \rangle = \langle N_+N_- \rangle$, 
$\langle N_{++} \rangle = \langle N_+(N_+ - 1) \rangle$ etc., and 
simplified as
\begin{eqnarray}
\label{eq:16}
\Sigma[N_+, N_-] - 1 &=&
 \frac{\langle N_+ \rangle \langle N_- \rangle}{\langle N_+\rangle + 
                             \langle N_- \rangle}  \\
\nonumber 
                     && \times 
 \left\{\frac{\langle N_{++} \rangle}{\langle N_+\rangle^2} +
        \frac{\langle N_{--} \rangle}{\langle N_- \rangle^2} - 
      2 \frac{\langle N_+N_- \rangle}{\langle N_+\rangle 
                                      \langle N_-\rangle}\right\} \\
\nonumber
                &=& \frac{1}{\langle N_\text{ch}\rangle}\bigg\{
         \langle N_{++}\rangle + \langle N_{--}\rangle - 
                                2\langle N_{+-}\rangle\bigg\} \ .
\end{eqnarray}
Note that combining Eqs.(\ref{eq:15}) and (\ref{eq:16}), 
Eq.(\ref{eq:13}) is obtained.

An important property of a strongly intensive quantity is its 
independence on volume and volume fluctuations. It is worth noting
that $\Sigma[N_+, N_-]$ is equal to unity in case of independent
emission of particles in the so-called independent particle model
\cite{Gazdzicki:2013ana,Wu:2021vix}.

The $D$ and $\Sigma$ variables measure both net-charge fluctuations 
and correlations. Some apparent sources of correlations are resonance 
decays in final state and production of (mini)jets, which are not 
thermalized in QGP. Moreover, the trivial correlations of particles 
caused by charge conservation have to be accounted for, given that 
the predictions for the magnitudes of fluctuations were derived using 
a grand canonical ensemble where charge is conserved on average. The 
correction for global charge conservation of $\nu_{(+-,dyn)}$ 
proposed in \cite{Pruneau:2002yf} reads
\begin{equation}
\label{eq:Cor_V}
\nu_{+-,\text{dyn}}^\text{corr} =  \nu_{+-,\text{dyn}} + 
                           \frac{4}{\langle N_{tot} \rangle}\ ,
\end{equation}
where $\langle N_{tot} \rangle$ is the total charged multiplicity in 
full phase space. Another approach to correct the $D$ magnitude is 
derived in \cite{BJK00},
\begin{equation}
\label{eq:Cor_B}
D^{\text{corr}} = (\nu_{+-,\text{dyn}}\langle N_{ch}\rangle + 4)/
                  (C_{\mu}C_{\eta}) \ ,
\end{equation}
where $C_{\mu} =\langle N_{+}\rangle^{2} /\langle N_{-}\rangle^{2}$ 
and $C_{\eta} =1- \langle N_{ch}\rangle /\langle N_{tot}\rangle$.

\section{Numerical results for $D$ and $\Sigma$ variables}
\label{sec_4}

\begin{figure}[htpb]
\includegraphics[width=0.49\textwidth]{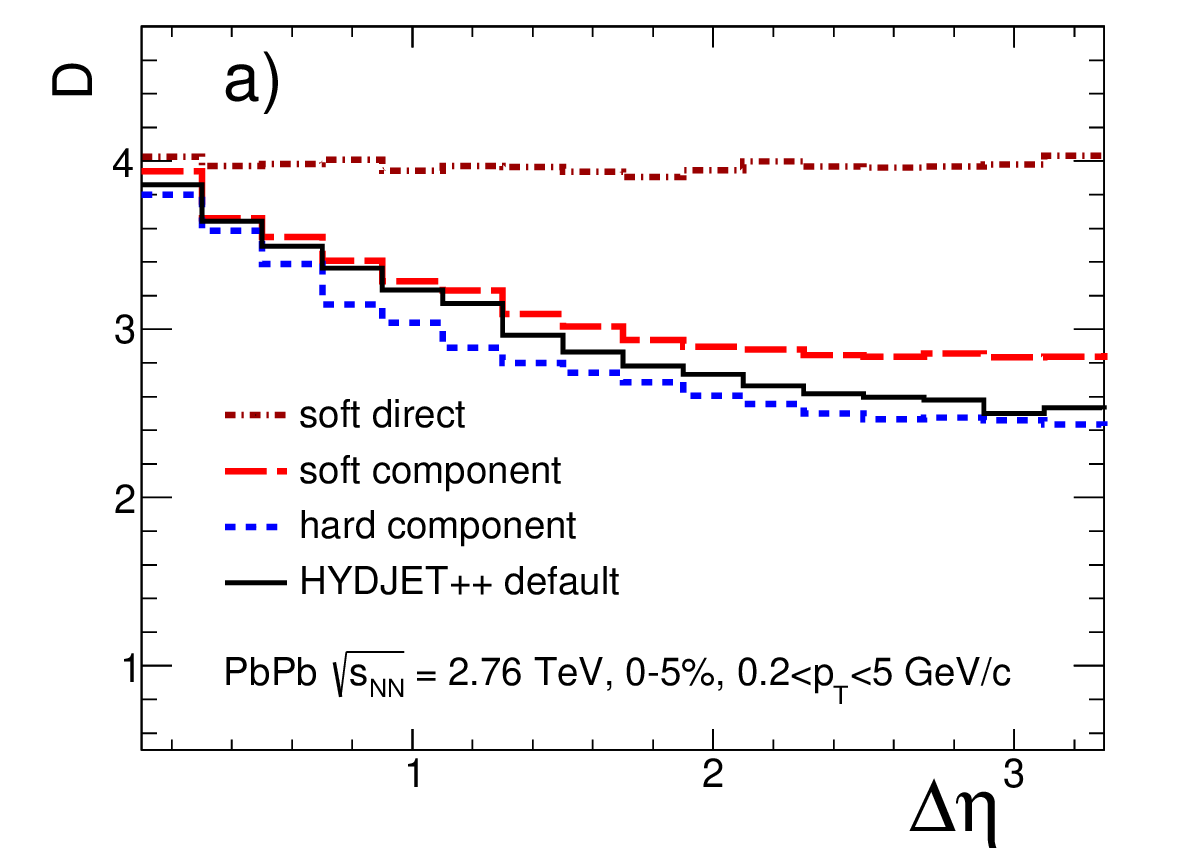}
\includegraphics[width=0.49\textwidth]{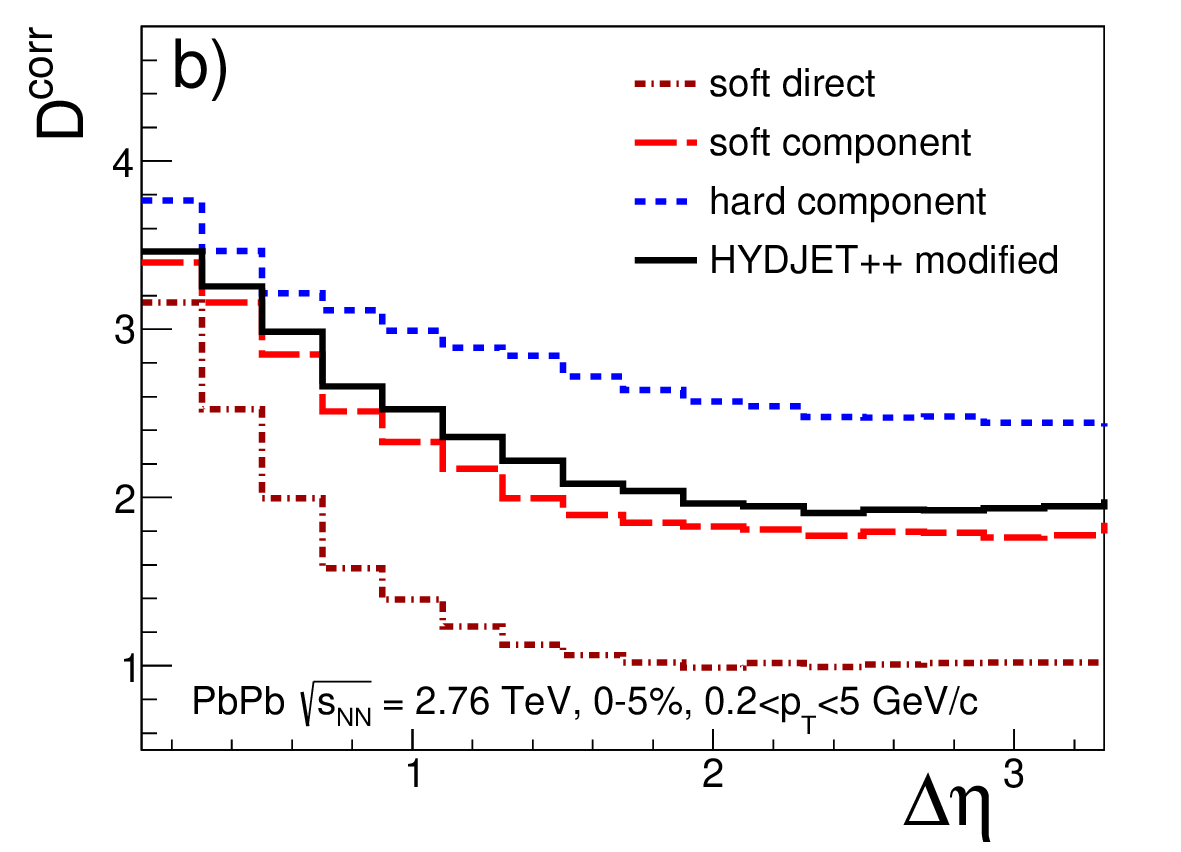}
\caption{ (color online)
$D$ and $D^{corr}$ represented as a function of the $\Delta\eta$ 
window for the \hydjet model-generated \PbPb collisions at 
$\sqrts = 2.76$~TeV with centrality of $0-5$\% for 
each component: soft (red dashed line), hard (blue dotted line), 
for direct soft hadrons (brown dash-dotted line) together with the 
resulting total value (full black line); {\bf (a)} \hydjet 
calculations in default mode; {\bf (b)} modified model calculations.}
\label{Fig_1}
\end{figure}

The soft part of the \hydjet model represents a grand canonical 
ensemble. Thus, the correction for global charge conservation is not 
needed. The primordial, or direct, hadron production in the soft 
component corresponds to independent particle emission and is 
expected to be similar to the $D$ value predicted for a hadron gas, 
i.e., $D=4$ or $\Sigma=1$. This is observed in Fig.~\ref{Fig_1}(a) 
for the directly produced hadrons. The correlations of charged 
hadrons in jets are stronger than those for resonances, and typically 
result in even smaller values of $D$ ($\Sigma$). Note, however, that 
the hard component of \hydjet is corrected for global charge 
conservation during the jet fragmentation process 
(see Eq.(\ref{eq:Cor_V})). The correction is more dramatic for larger 
phase space.
We checked that the hard component of the $D$-measure exhibits low 
sensitivity to partial thermalization owing to jet quenching and to 
the electrical charge of the initial parton, gluon or quark.

Some models experience difficulties to describe the data. Model 
calculations usually provide higher values of $D$ ($\Sigma$). 
Moreover, experimental results exhibit centrality dependence not 
reproduced by the models. The low value of $D$ may arise either from 
the small net-charged fluctuations due to more uniformly distributed 
charges in a QGP or from the strong $(+-)$ correlations. The 
modifications in the \hydjet model were introduced to describe the 
width of the experimental BF. In particular, the $(+-)$ pair 
production of direct hadrons was employed in the soft component, 
implying strong correlations (with short correlation length) among 
unlike-sign charged hadrons. Such correlations sufficiently reduce 
the values of $D$, as shown in Fig.~\ref{Fig_1}(b). Note that, in 
this case, the resonance decays dilute the correlations, thus 
increasing the $D$ magnitude, in contrast to the situation with 
hadron gas, where resonance decays decrease the $D$ magnitude. This 
effect of resonance decays may also be valid for the case of QGP 
fluctuations. Therefore, the magnitude of fluctuations for the QGP 
may be increased by
\begin{itemize}
\item resonance decays in the final state
\item contribution of non-thermalized, or partially thermalized, 
hard mini-jets, which are still present within the considered low 
$p_{T}$ interval, $0.2 < p_{\rm T} < 5$~GeV/$c$.
\end{itemize}
To date, the corrected values of $D$ experimentally measured are 
larger than the expectations for QGP.

\begin{figure}[htpb]
\includegraphics[width=0.49\textwidth]{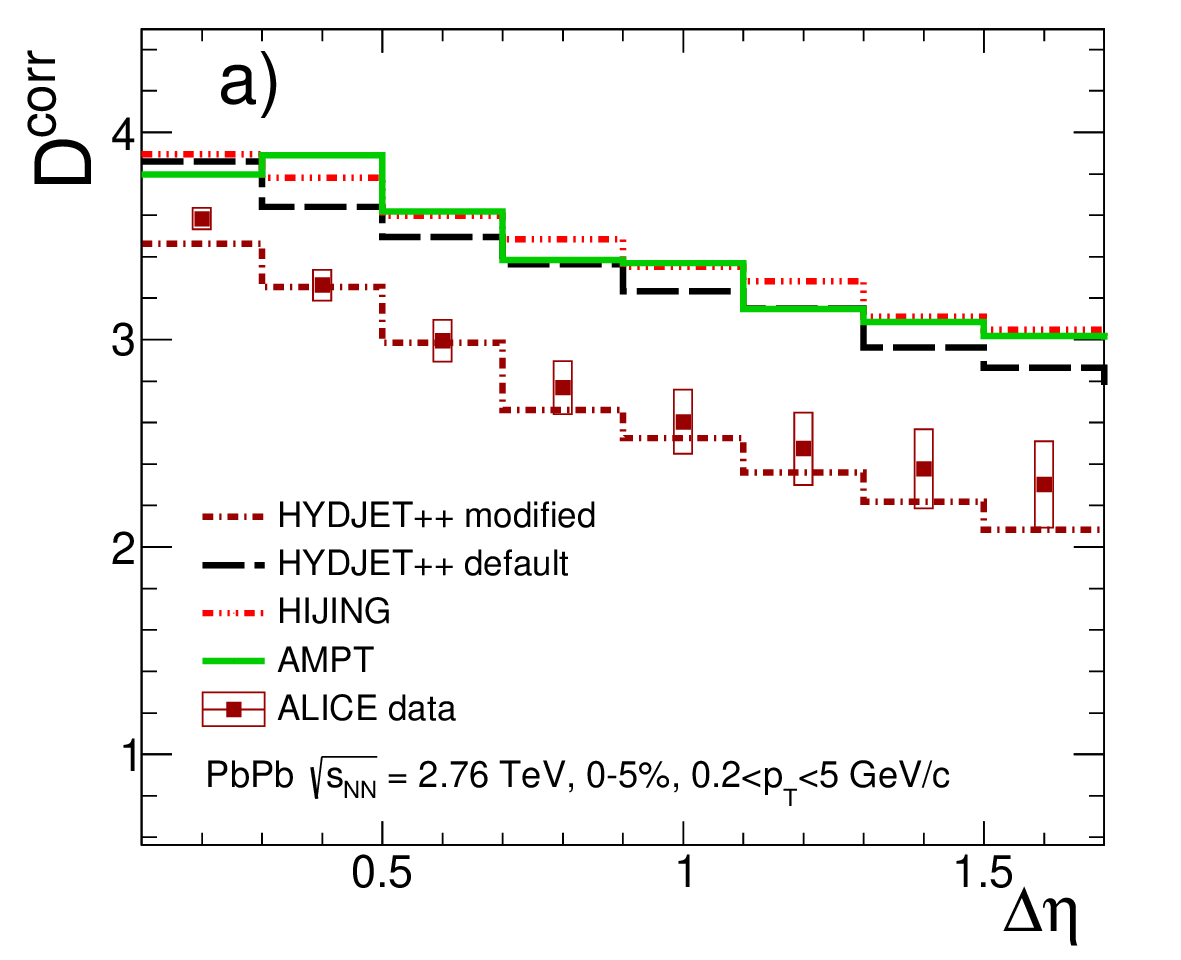}
\includegraphics[width=0.49\textwidth]{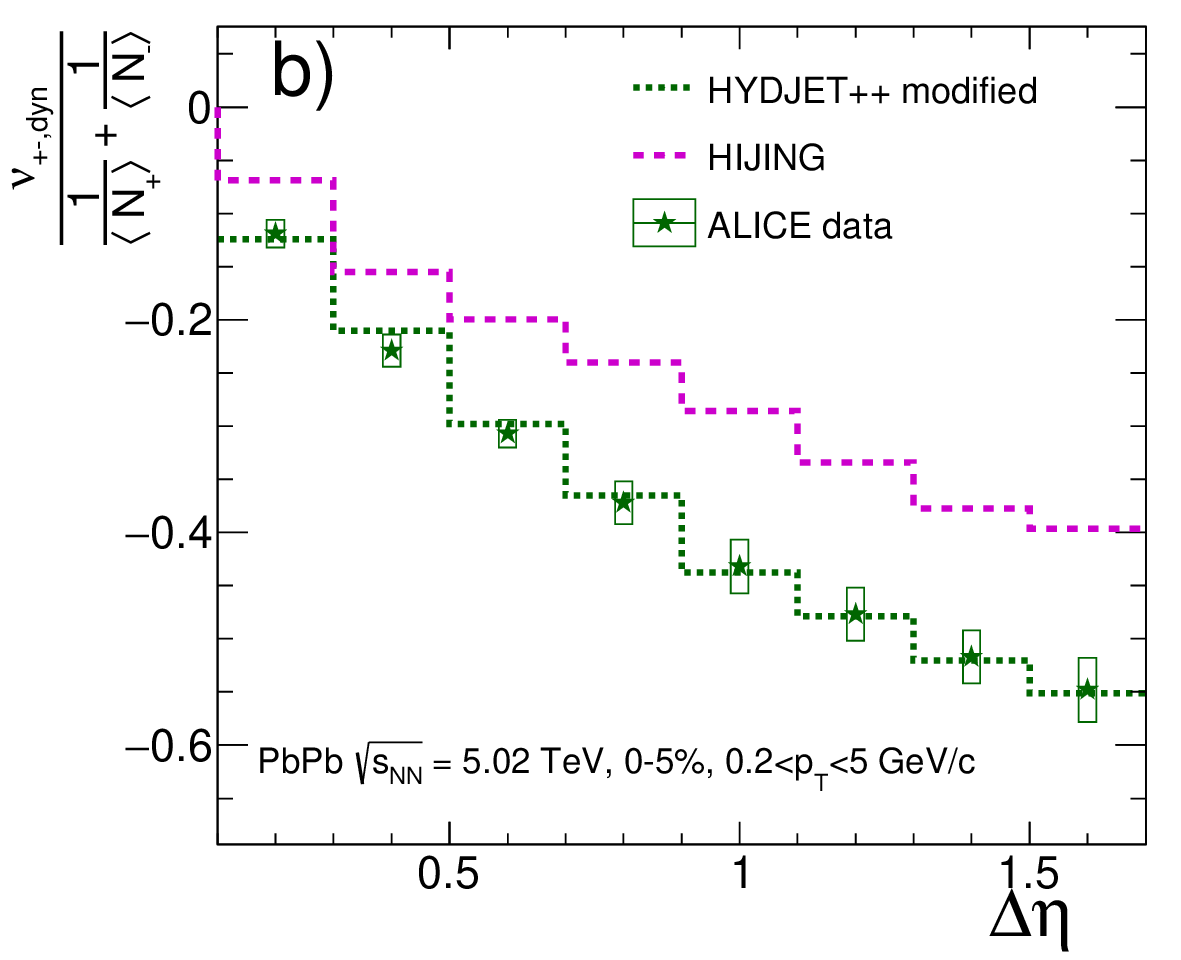}
\caption{ (color online)
(a): $D$ and $D^\text{corr}$ represented as a function of the 
$\Delta\eta$ window for the \hydjet (dashed histogram), \hydjet 
modified (dash-dotted), \ampt (full), and \hijing (dotted-dash) 
model-generated \PbPb collisions at 2.76~TeV with centrality of 
0-5\%, in comparison with data (squares). These data were extracted 
from \cite{ALICE:2012xn};
(b): 
$\ds \nu_{+-,\text{dyn}}\,\bigg/ \left( \frac{1}
{\langle N_+\rangle} + \frac{1}{\langle N_-\rangle} \right) $ as 
a function of $\Delta\eta$ for \hydjet modified (dotted histogram) 
and \hijing (dashed) model-generated \PbPb collisions at 5.02 TeV with 
centrality of 0-5\% in comparison with data (stars) extracted from 
\cite{Khan:2022ndt}.}
\label{Fig_2}
\end{figure}

Collisions of lead beams at the LHC give rise to events with the 
highest multiplicity of produced particles. A comparison of the 
\ampt, \hijing, and \hydjet models with the experimental values of
$D^{\text{corr}}(\Delta\eta)$ in central 0-5\% Pb+Pb collisions at 
$\sqrts = 2.76$~TeV \cite{ALICE:2012xn} is depicted in 
Fig.~\ref{Fig_2}(a). The $0.2 < p_{\rm T} < 5$~GeV/$c$ transverse 
momentum range was used in the calculations, similar to the 
experimental data. Note that the \hijing and modified \hydjet models 
are corrected for global charge conservation, while the \ampt and 
\hydjet default models are not corrected, given that the global 
electric charge is not conserved, in general, in each individual 
event. For instance, as pointed out in \cite{ZWL14_app,ZWL21}, 
violation of the electric charge conservation in the \ampt model 
occurs because of two reasons. One reason is that, in some many-body 
decays of resonances and many hadronic interactions, the electric 
charge of each hadron in the final state is set randomly 
\cite{ZWL14_app}. The second reason is that the hadronic cascade in 
the \art model treats only charged kaons as explicit particles,
discarding the neutral ones.

The corrected values of $D^{\text{corr}}$ were obtained as averaged 
values provided by Eqs.(\ref{eq:Cor_V}) and (\ref{eq:Cor_B}), as 
performed for experimental data processing. A comparison of model 
results with experimental $\Sigma -1$ values, expressed as 
$\ds \nu_{+-,\text{dyn}}\,\bigg/ \left( \frac{1}
{\langle N_+\rangle} + \frac{1}{\langle N_-\rangle} \right)$ 
for central 0-5\% Pb+Pb collisions at $\sqrts = 5.02$~TeV 
\cite{Khan:2022ndt}, is presented in Fig.~\ref{Fig_2}(b).

\begin{figure}[htpb]
\includegraphics[width=0.49\textwidth]{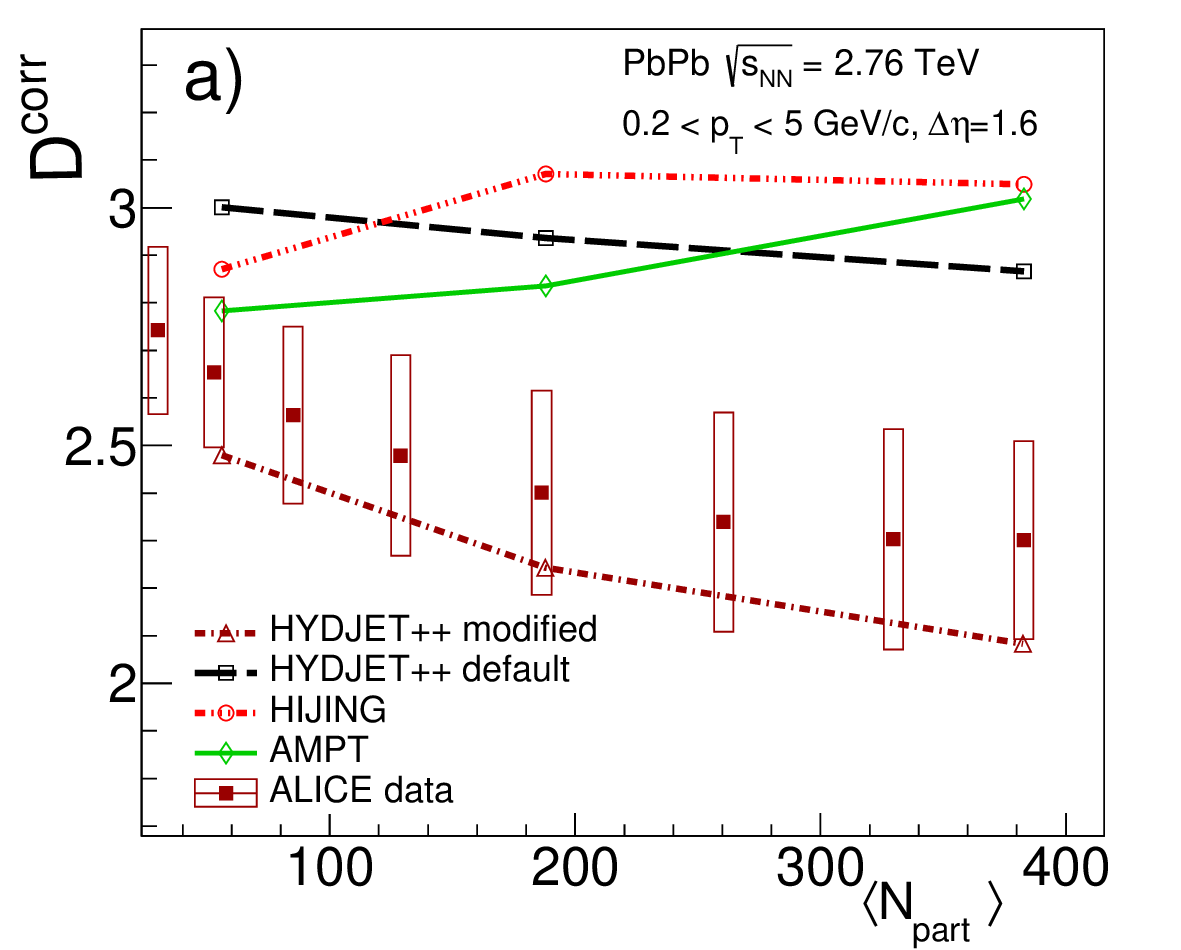}
\includegraphics[width=0.49\textwidth]{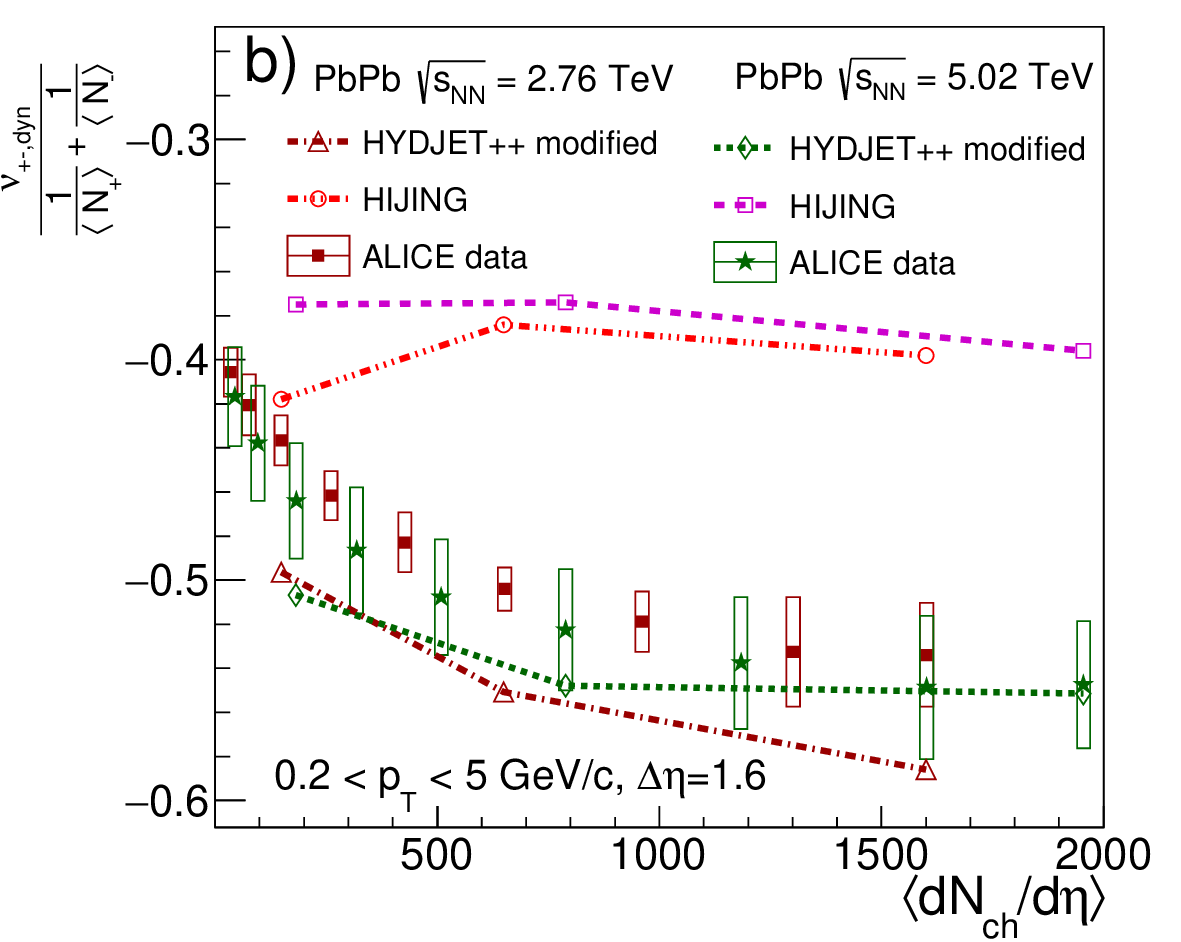}
\caption{ (color online)
(a): $D$ and $D^\text{corr}$ represented as a function of $N_{part}$ 
for the \hydjet (open squares), \hydjet modified (open triangles), 
\ampt (open diamonds), and \hijing (open circles) model-generated 
\PbPb collisions at 2.76 TeV in comparison with ALICE data (full 
squares) extracted from \cite{ALICE:2012xn}; the lines are drawn to 
facilitate visualization;
(b): 
$\ds \nu_{+-,\text{dyn}}\,\bigg/ \left( \frac{1}
{\langle N_+\rangle} + \frac{1}{\langle N_-\rangle} \right) $ 
represented as a function of $\Delta\eta$ for the \hydjet modified 
(open triangles and open diamonds) and \hijing (open circles and 
open squares) model-generated \PbPb collisions at 2.76~TeV and 
5.02~TeV, respectively, in comparison with ALICE data (full squares 
and full stars) extracted from \cite{ALICE:2012xn} and 
\cite{Khan:2022ndt}; again, the lines are drawn to facilitate
visualization.}
\label{Fig_3}
\end{figure}

A comparison with experimental centrality dependence of measured 
fluctuation values ($D$ and $\Sigma-1$) for Pb+Pb collisions at 
$\sqrts = 2.76$~TeV \cite{ALICE:2012xn} and $\sqrts = 5.02$~TeV 
\cite{Khan:2022ndt} is displayed in Fig.~\ref{Fig_3}. The 
centrality dependence in ALICE data is represented as a function of 
$\langle N_{part}\rangle$ and $\langle dN_{ch}/d\eta\rangle$. The 
model results correspond to centralities of 0-5\%, 20-30\%, and 
50-60\%, respectively. The pseudorapidity windows are chosen to be 
symmetric around $\eta = 0$. Note that the results for the
$\ds \nu_{+-,\text{dyn}}\,\bigg/ \left( \frac{1}
{\langle N_+\rangle} + \frac{1}{\langle N_-\rangle} \right)$ 
variable for both \ampt and default \hydjet models are not shown here 
because the data are not corrected for global charge conservation.
Note also that the \hydjet model with modifications qualitatively 
describes the data. The strength of pair correlations was adjusted 
to describe the BF widths, whereas the $D$-measure is connected to 
the integral of the balance function, where both the width and 
amplitude of BF are important. To quantitatively describe the data, 
the strength of the pair correlations $weta \equiv 
\sigma_{\eta}$ (see Eq.(\ref{gauss})) was reduced as listed in 
Table~\ref{table1}. 

\begin{table}
\caption{Values of previous and newly tuned parameter {\it weta}.}    
\begin{center}
 \begin{tabular}{l|c|c|c|}
   \hline \hline 
     Centrality & 0-5\% & 20-30\% & 50-60\% \\
      \hline

     \hydjet modified weta & 0.35 & 0.5 & 1\\

     \hydjet new tune weta & 0.5 & 0.7 & 1.1\\

   \hline \hline
 \end{tabular}
\end{center}
\label{table1}
\end{table}

\begin{figure}[htpb]
\includegraphics[width=0.49\textwidth]{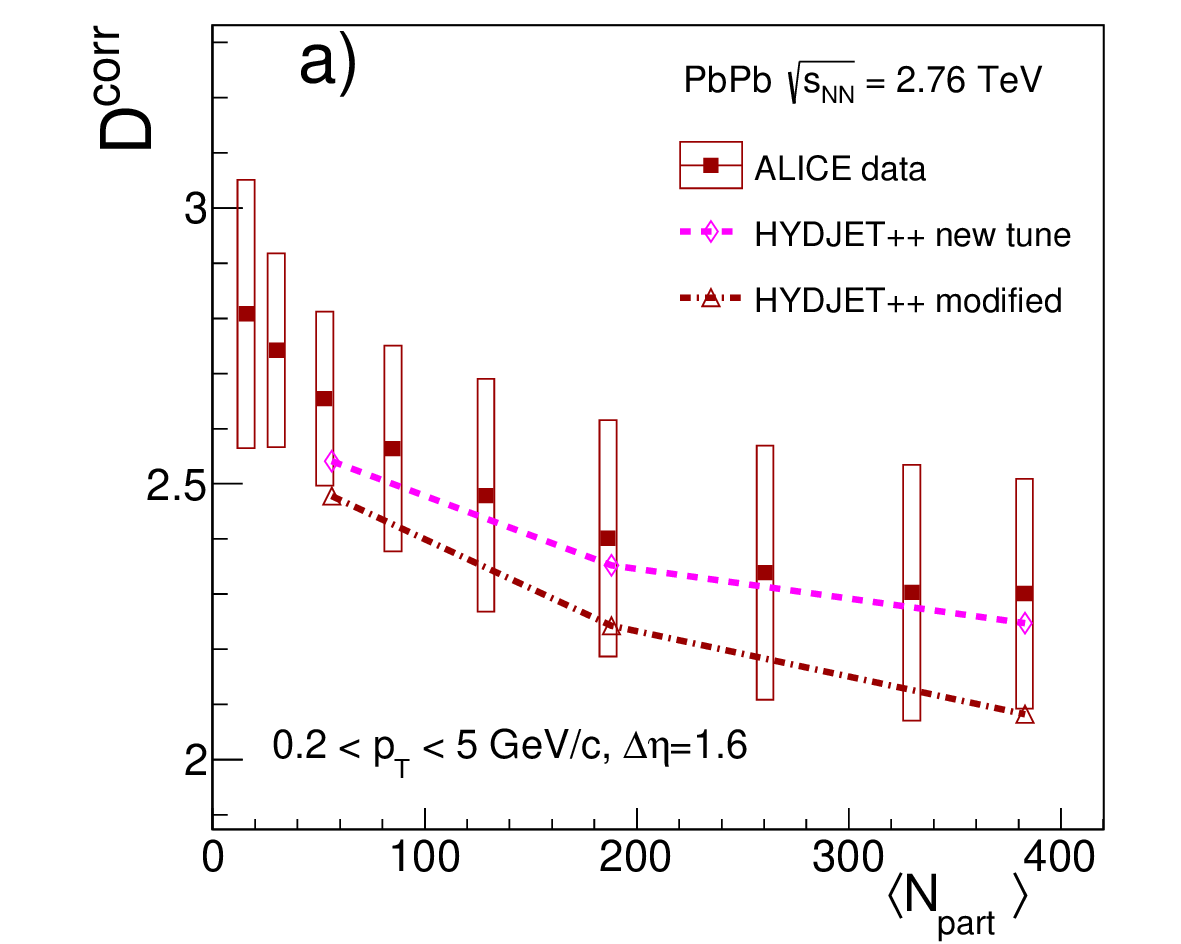}
\includegraphics[width=0.49\textwidth]{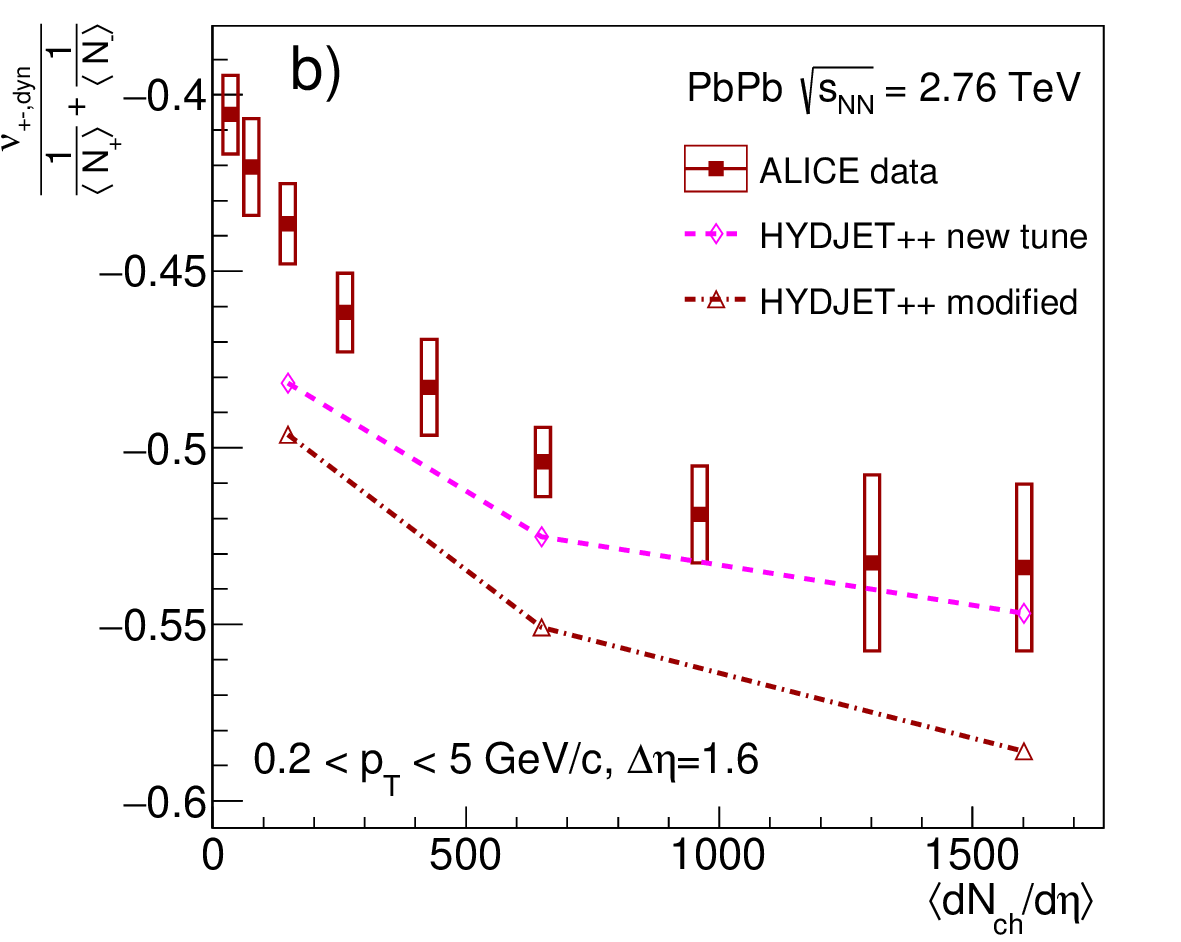}
\caption{(color online) Same as Fig.~\ref{Fig_3}, but with reduced 
correlation strength in the \hydjet model. Please, refer to the main 
text for further details.}
\label{Fig_4}
\end{figure}

Results of calculations made with previous and newly tuned parameters 
are displayed in Fig.~\ref{Fig_4}. The \hydjet model adequatly 
reproduces experimental data both for $D^{\text{corr}}$ and 
$\Sigma[N_+, N_-] - 1$.
Therefore, it can be concluded that the main shortcoming of various
models when it comes to reproducing data is either the use of a grand 
canonical ensemble to describe multiparticle production, which is the
case of macroscopic statistical models (e.g., \hydjet default), or 
the random distribution of electric charge for secondary hadrons in 
microscopic transport models (e.g., \ampt). To solve this problem, it 
is necessary to take into account the pair production of hadrons of 
opposite sign (\hydjet modified).

We also checked that the new parametrization does not affect the 
widths of the balance functions of charge particles studied using 
the \hydjet model in \cite{Chernyshov:2022oik}.
Recall that the charge balance function is defined as
\begin{eqnarray}
\displaystyle
\label{bf}
B(\Delta \eta) &=& \frac{1}{2}  \Big[\frac{\langle N_{+-}(\Delta \eta)
   \rangle - \langle N_{++}(\Delta \eta)\rangle }{\langle N_{+}
   \rangle } \nonumber \\
&+&\frac{\langle N_{-+}(\Delta \eta)\rangle - \langle N_{--}(\Delta \eta)
   \rangle }{\langle N_{-} \rangle }\Big]\ ,
\end{eqnarray}
where $\langle N_{+-}(\Delta \eta) \rangle$ is the average number of
unlike-charge pairs with particles separated by the relative
pseudorapidity $\Delta \eta =\eta_{+}-\eta_{-}$, and this is also 
the case for $\langle N_{-+}(\Delta \eta) \rangle$, 
$\langle N_{++}(\Delta \eta) \rangle $, and 
$\langle N_{--}(\Delta \eta) \rangle $.

\begin{figure}[htpb]
\includegraphics[width=0.47\textwidth]{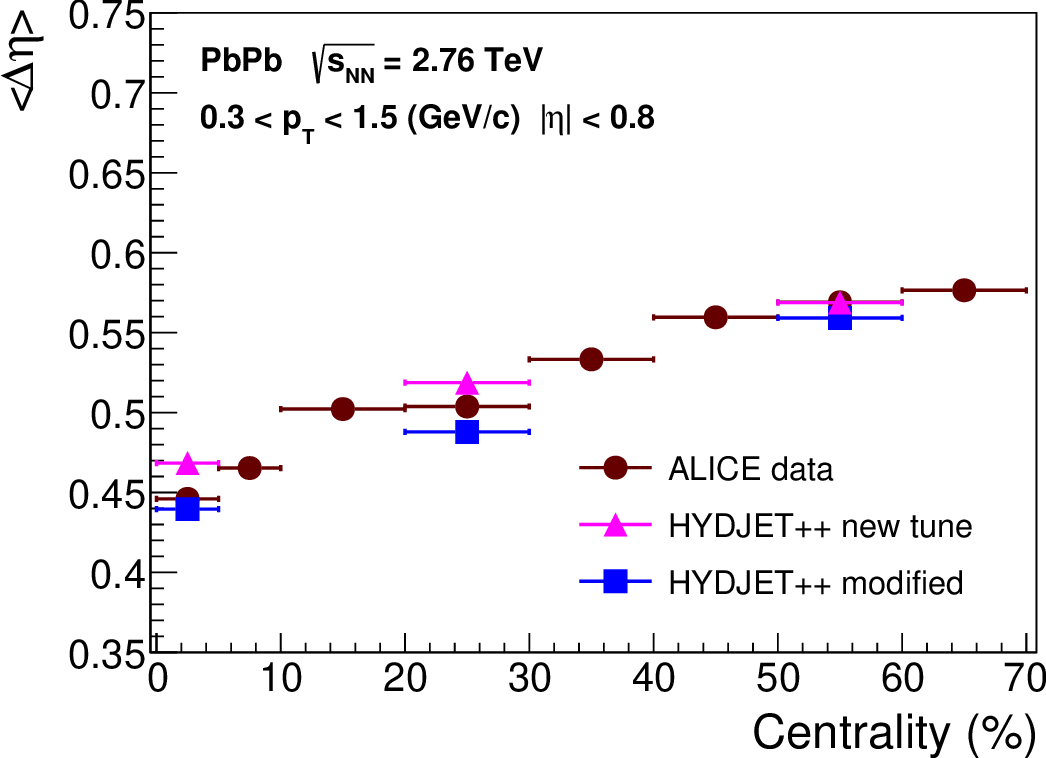}
\caption{ (color online)
Centrality dependence of the width of the balance function of charged
hadrons in \PbPb collisions at $\sqrts = 2.76$~TeV for the 
correlations studied in terms of relative pseudorapidity $\Delta 
\eta$. The model calculations with modified (squares) and newly tuned 
(triangles) versions of the \hydjet model are compared with the ALICE 
data (circles) extracted from \cite{alice_plb_723}. }
\label{Fig_5}
\end{figure}

Variations of the width of the balance function of charged hadrons 
for the correlations studied as functions of relative pseudorapidity 
$\Delta \eta$ with centrality in \PbPb collisions at $\sqrts = 
2.76$~TeV are shown in Fig.~\ref{Fig_5}. Note that the deviations of 
model calculations with newly tuned parameters from those presented 
in \cite{Chernyshov:2022oik} lie within a 7\% accuracy limit.

\vspace{0.2cm}

\begin{figure}[htpb]
\includegraphics[width=0.48\textwidth]{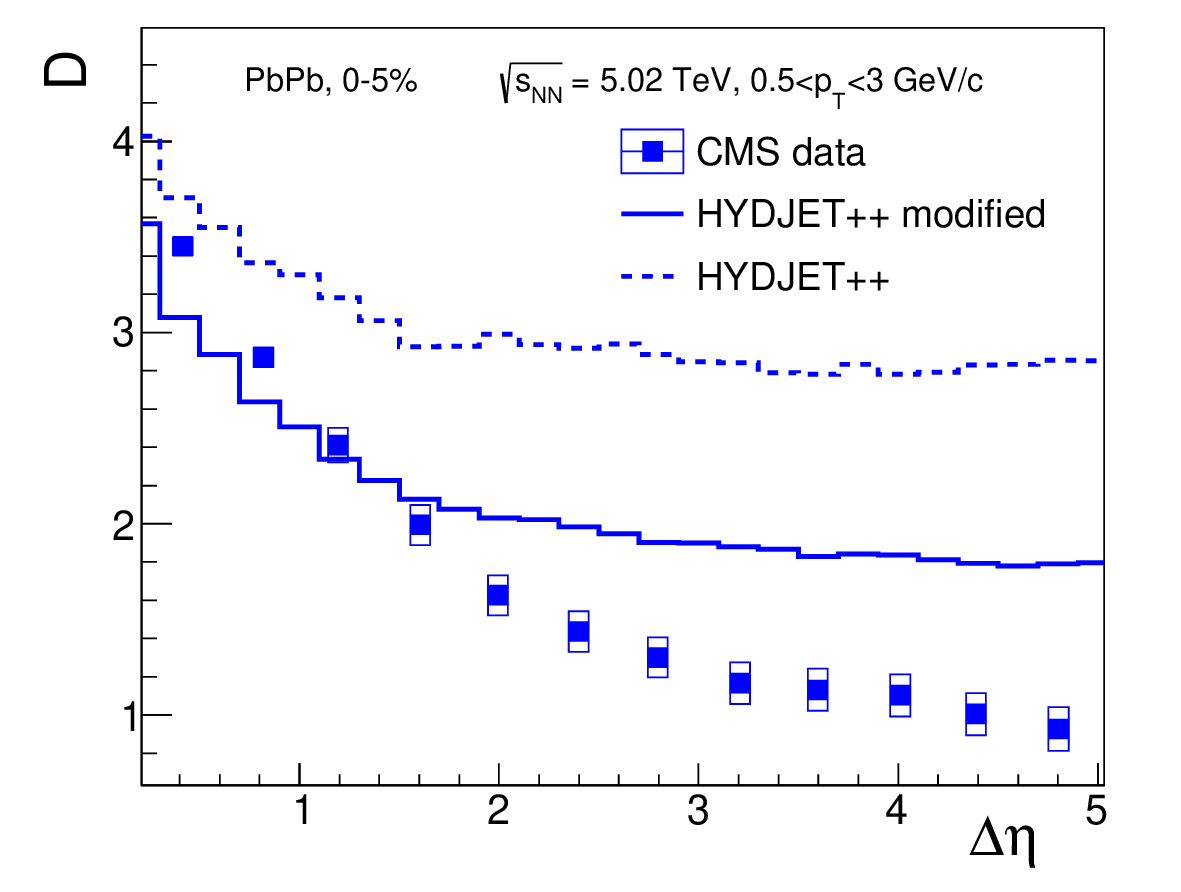}
\caption{ (color online)
$D$ as a function of $\Delta \eta$ in \PbPb central ($0 - 5\%$) 
collisions at $\sqrts = 5.02$~TeV. The squares represent CMS data 
extracted from \cite{hin-22-005}; \hydjet calculations for the 
default and modified versions are shown by dashed and solid 
histograms, respectively. }
\label{Fig_6}
\end{figure}

Recently \cite{hin-22-005}, the CMS Collaboration presented  
preliminary data on the net-charge fluctuations in \PbPb collisions at 
$\sqrts = 5.02$~TeV for a wide range of relative pseudorapidity up to 
$\Delta \eta = 4.8$. The analysis was performed in terms of the
variable $D$. The default \hydjet model fails to accurately replicate 
the data, as shown in Fig.~\ref{Fig_6}. In contrast, the modified version 
of \hydjet reproduces the experimentally measured values up to 
$\Delta \eta = 1.5$ but overestimates them at larger pseudorapidities. 
The discrepancy may be explained by a possible shift of the strength of 
pair correlations $\sigma_{\eta}$ at large rapidities. Note also that 
the CMS data are not corrected for the effect of global charge 
conservation, which is more prominent at large values of $\Delta \eta$. 
This effect might reveal itself differently in data and in model 
calculations. In other words, for a fixed value of $\sigma_{\eta}$, 
the effect of charge fluctuations due to directly produced hadrons 
decreases with increasing $\Delta \eta$, given that an increasingly 
smaller fraction of charged hadrons from each generated pair will not 
fall within the considered range of pseudorapidity. It is this 
exclusion from the observation zone that provides the net-charge 
fluctuations of direct hadrons. This important question deserves 
further investigation.

\section{Discussion and Conclusions}
\label{sec_5}

Fluctuations of the net electric charge of particles are studied in 
\PbPb collisions at LHC energies of $\sqrts = 2.76$~TeV and 5.02~TeV  
within the framework of several models designed for description of 
nucleus-nucleus interactions at (ultra)relativistic energies.
The following conclusions can be drawn from our study.

The values of quantitative measures of these fluctuations, namely 
$D$ and $\Sigma$, obtained from the \hydjet, \ampt, and \hijing 
models, exceed the corresponding experimental values and are 
more consistent with the fluctuations corresponding to those of a 
hadronic gas. The discrepancy is attributed to the treatment of 
multiparticle production within the grand canonical ensemble 
approach rather than the canonical one, which accounts for the 
explicit charge conservation in each individual event. 

A modified version of the \hydjet model with pair-production of
oppositely charged hadrons on the freeze-out hypersurface, in 
contrast to independent generation of hadron species in the
``default" version of the model, properly describes the data. 
This is achieved by introducing a new model parameter, namely 
charge correlation length at the formation stage of direct 
hadrons. This effectively considers the mechanisms of occurrence 
of the charge correlations. In this case, the charge correlation 
length decreases when going from the peripheral collisions to the 
central ones. This behavior of the parameters $D$ and $\Sigma$ 
corresponds to fluctuations attributed to those of the quark-gluon 
plasma. This modification and the corresponding parameters allow 
describing the width of the balance function of unlike-sign charged 
hadrons in lead-lead collisions at energies of $\sqrts = 2.76$~TeV 
and 5.02~TeV.

It has been established that the decay of resonances in the case 
of initially strong correlations of the directly produced hadrons 
with unlike charges, i.e., with a correlation length less than that 
for resonances, causes blurring of these correlations and, hence, 
increase of the parameter $D (\Sigma)$. In contrast, in case of 
absence of charge correlations between directly produced hadrons, 
which corresponds to an infinite correlation length, decays of 
resonances lead to smaller values of $D (\Sigma)$.

\section*{Acknowledgments}
{\it Fruitful discussions with L.V.~Bravina and
A.I.~Demyanov are highly appreciated.
A.Ch. acknowledges support from the BASIS Foundation under
grant No.23-2-10-2-1.}

\end{document}